\documentclass[aps,superscriptaddress,preprint]{revtex4}
\usepackage{amsfonts}
\usepackage{amsmath}
\usepackage{amssymb}
\usepackage{graphicx}

\begin{document}
\title{Experimental Observation of Quantum Holographic Imaging}
\author{Xin-Bing Song}
\affiliation{Department of Physics, Applied Optics Beijing Area
Major Laboratory, Beijing Normal University, Beijing 100875, China}
\author{De-Qin Xu}
\affiliation{Department of Physics, Applied Optics Beijing Area
Major Laboratory, Beijing Normal University, Beijing 100875, China}
\author{Hai-Bo Wang}
\affiliation{Department of Physics, Applied Optics Beijing Area
Major Laboratory, Beijing Normal University, Beijing 100875, China}
\author{Jun Xiong}
\affiliation{Department of Physics, Applied Optics Beijing Area
Major Laboratory, Beijing Normal University, Beijing 100875, China}
\author{Xiangdong Zhang}
\affiliation{Department of Physics, Applied Optics Beijing Area
Major Laboratory, Beijing Normal University, Beijing 100875, China}
\author{De-Zhong Cao}
\affiliation{Department of Physics, Yantai University, Yantai
264005, China}
\author{Kaige Wang}
\thanks{Author to whom correspondence should be addressed:wangkg@bnu.edu.cn}
\affiliation{Department of Physics, Applied
Optics Beijing Area Major Laboratory, Beijing Normal University,
Beijing 100875, China}
\pacs{PACS}

\begin{abstract}
We report the first experimental observation of quantum holographic
imaging with entangled photon pairs, generated in a spontaneous
parametric down-conversion process. The signal photons play both
roles of "object wave" and "reference wave" in holography but are
recorded by a point detector providing only encoding information,
while the idler photons travel freely and are locally manipulated
with spatial resolution. The holographic image is formed by the
two-photon correlation measurement, although both the signal and
idler beams are incoherent. According to the detection regime of the
signal photons, we analyze three types of quantum holography
schemes: point detection, coherent detection and bucket detection,
which can correspond to classical holography using a point source, a
plane-wave coherent source and a spatially incoherent source,
respectively. Our experiment demonstrates that the two-photon
holography in the point detection regime is equivalent to the
one-photon holography using a point source. Physically, the quantum
holography experiment verifies that a pair of non-commutable
physical quantities, the amplitude and phase components of the field
operator, can be nonlocally measured through two-photon
entanglement.
\end{abstract}
\maketitle

Holography, first proposed by Gabor in 1948\cite{Gabor}, is a
lensless imaging technique and capable of recording entire
information of an object. Different from usual photography, where
only the intensity of the optical field shining an object is
recorded, both the amplitude and phase of the field are recorded by
adding a reference field in holography. Hence holography requires a
coherent source with both better temporal and spatial coherence,
such as a laser beam, to perform the spatial interference between
the object wave and reference wave. A challenging question would be:
can holography be performed by other sources, which are not coherent
or even nonclassical? Recently, Zhang et al.\cite{zhang1,zhang2}
discovered that the spatial coherence is not necessary in the
holographic interference. Their schemes used an incoherent thermal
light source with an extended area, and the object wave and
reference wave are arranged to experience different diffraction
configurations. Different from coherent holography where the
holographic pattern is stationary, the interference pattern in the
incoherent regime fluctuates in time, but can be formed in the
statistical summation.

Spontaneous parametric down-conversion (SPDC) process in a nonlinear
crystal may generate a nonclassical light source - the two-photon
quantum entangled state, which is very close to the
Einstein-Podolsky-Rosen (EPR) state\cite{ein1}. The two
down-converted beams in SPDC are incoherent, but the coherence can
be revived in the two-photon correlation. In the pioneer theoretical
work of two-photon optics, Belinskii and Klyshko\cite{kly} predicted
three spooky schemes: two-photon diffraction, two-photon holography,
and two-photon transformation of two-dimensional images. The first
and last schemes have been demonstrated in the experiments, known as
ghost interference\cite{shih1} and ghost imaging\cite{shih2},
respectively. These experiments were regarded as close to the
original \emph{gedankenexperiment} of EPR paradox, since the
position or momentum information detected by one photon can be
nonlocally transferred to the other photon. However, to our best
knowledge, the two-photon holography has not been tested
experimentally so far.

In 2001, Abouraddy et al \cite{teich1} proposed a theoretical scheme
of quantum holography using a two-photon entangled source. In their
scheme, one photon of the entangled photon pair illuminates the
remote object and then is collected by a bucket detector while the
other is locally manipulated providing conventional spatial
resolution. Since quantum entanglement behaves as ``spooky actions
at a distance'' (in Einstein's word)\cite{ein2}, the holographic
information of the remote object can be recorded by the coincidence
measurement of the two photons. They claimed that quantum holography
is particularly suitable for imaging of a hidden object or an object
in a confined space where the conventional imaging is impossible.
Later, they realized the entangled-photon ghost imaging experiment
with a pure phase object, but not holographic imaging\cite{teich2}.

In this paper, we report the first experimental observation of
holographic imaging using a two-photon entangled source. In our
detailed theoretical analysis, we find that the quantum holography
scheme in terms of bucket detection proposed by Abouraddy et al
\cite{teich1} is restricted in the experimental performance. We
compare two different detection regimes in the two-photon quantum
holography, the bucket detection and the point detection, both of
which record the encoding information of the photon shining the
object. As a matter of fact, quantum holography fails if the bucket
detection is applied to the holographic interference where the two
interfered waves experience the same diffraction length. However,
the point detection regime is adequate for the equal-path
holographic interference, which is employed in our experiment.

We first recast classical holography with a simple in-line
interferometric scheme, as sketched in Fig. 1(a). The beam from a
source is divided into two daughter beams by a beamsplitter: one
illuminates an object while the other travels freely, called object
wave and reference wave, respectively. The two waves interfere at
the recording material to form a hologram. Let $x$ and $x_0$ be the
transverse positions across the beam, $E_o(x)$ and $E_r(x)$ are the
fields in the recording plane for the object wave and reference
wave, respectively, and they satisfy
\begin{equation}\label{1}
E_j(x)=\int h_j(x,x_0)E_0(x_0)dx_0,  (j=o,r)
\end{equation}
where $E_0(x_0)$ is the field distribution in the source plane.
$h_j(x,x_0)$ stands for the impulse response function (IRF) for path
$j=o,r$. Under the paraxial approximation, the IRF of the object
wave and reference wave are written as
\begin{subequations}
\label{2}
\begin{eqnarray}
h_o(x,x_0)&=&\frac{k\exp(ikz_{o})}{i2\pi\sqrt{z_{o1}z_{o2}}}
\int dx' T(x')\exp \left[\frac{ik(x_0-x')^2}{2z_{o1}}+\frac{ik(x'-x)^2}{2z_{o2}}\right],\label{2a} \\
h_r(x,x_0)&=& H(x,x_0,z_r)\equiv\sqrt{\frac{k}{i2\pi
z_r}}\exp(ikz_r)\exp\left[\frac{ik(x-x_0)^2}{2z_r}\right],\label{2b}
\end{eqnarray}
\end{subequations}
respectively. $k$ is the wave number of the beam. $z_{o1}$ and
$z_{o2}$ are the distances from object to source and recording
plane, respectively, and $z_{o}=z_{o1}+z_{o2}$; $z_r$ is the
diffraction length for the reference wave. For the convenience of
theoretical treatment, we assume an transmissive object described by
Function $T(x')$.

In holography, the object wave is usually much weaker than the
reference wave. So the holographic pattern in the recording plane is
dominated by the interference term $\langle
E^{\ast}_r(x)E_o(x)\rangle$. When the temporal coherence condition
is satisfied, that is $|z_r-z_o|$ is much less than the longitudinal
coherence length of the beam, one arrives
\begin{equation}\label{3}
\langle E^{\ast}_r(x)E_o(x)\rangle=\int dx'_0dx_0
h^{\ast}_r(x,x'_0)h_o(x,x_0)\langle E^{\ast}_0(x'_0)E_0(x_0)\rangle.
\end{equation}

We consider three types of light sources in the spatial
interference. The first source is a plane-wave coherent field, for
which $\langle E^{\ast}_0(x'_0)E_0(x_0)\rangle=\alpha^{\ast}\alpha$
is independent of transverse positions. Hence the first-order field
correlation function can be factorized to be
\begin{eqnarray}
&&\langle E^{\ast}_r(x)E_o(x)\rangle=E^{\ast}_r(x)E_o(x) \nonumber\\
&&=|\alpha|^2\sqrt{\frac{k}{i2\pi z_{o2}}}\exp[ik(z_o-z_r)] \int
dx'T(x')\exp[ik(x-x')^2/(2z_{o2})],\label{4}
\end{eqnarray}
which records the holographic information of the object $T(x)$.

The second one is a thermal light source, shielded by a pinhole to
improve the spatial coherence. This type of source was originally
used in the first holography experiment\cite{Gabor}. In this case,
Eq.(\ref{1}) is reduced to $E_j(x)=h_j(x,x_0)E_0(x_0)\Delta x_0$,
where $x_0$ and $\Delta x_0$ are the position and width of the
pinhole, respectively. For simplicity, we assume $x_0=0$ and define
$E_0(0)\Delta x_0\equiv \beta$. Again, the first-order field
correlation function is factorized to be
\begin{eqnarray}
&&E^{\ast}_r(x)E_o(x)
=\left(\frac{k}{2\pi}\right)^{3/2}\frac{|\beta|^2\exp\left[ik(z_o-z_r)\right]}{\sqrt{iz_{r}z_{o1}z_{o2}}}
\exp\left[\frac{ik(z_r-z_o)x^2}{2z_rz_o}\right]\nonumber\\
&&\times\int
dx'T(x')\exp\left[\frac{ik}{2Z}\left(x'-\frac{x}{1+z_{o2}/z_{o1}}\right)^2\right],\label{5}
\end{eqnarray}
where the effective diffraction length is $Z=z_{o1}z_{o2}/z_o$.
Since the longitudinal coherence length of true thermal light is
very short, one must choose the equal-path configuration, i.e.
$z_r=z_o$. So the quadratic phase factor term outside the
integration disappears, and Eq. (\ref{5}) has the similar form as
Eq. (\ref{4}). Especially when the object is far from the source,
i.e. $z_{o1}>>z_{o2}$, the two equations become the same.

The last one is an incoherent thermal light source with an extended
area, which satisfies $\langle
E^{\ast}_0(x'_0)E_0(x_0)\rangle=I_0\delta(x'_0-x_0)$. As has
indicated above, this type of source is capable of performing
incoherent interference under the certain
conditions\cite{zhang1,zhang2}. Using Eq. (\ref{3}) we obtain
\begin{eqnarray}
&&\langle E^{\ast}_r(x)E_o(x)\rangle=I_0\int dx_0
h^{\ast}_r(x,x_0)h_o(x,x_0)\nonumber\\
&&=\frac{kI_0\exp[ik(z_o-z_r)]}{2\pi\sqrt{z_{o2}(z_r-z_{o1})}} \int
dx'T(x')\exp \left[\frac{ik(x'-x)^2}{2Z'}\right],\label{6}
\end{eqnarray}
where the effective diffraction length is
$Z'=z_{o2}(z_{r}-z_{o1})/(z_r-z_o)$. Apparently, the scheme fails
under the equal-path case because $Z'\rightarrow\infty$. However,
the poor temporal coherence of a true thermal light source requires
the equal-path condition in the interferometry. Hence this conflict
results in conventional opinion that a true thermal light source
with extended area is not appropriate for holographic
interferometry. Recent experiment demonstrated that a pseudo-thermal
light source associated with a laser having a long coherence time
can accomplish this incoherent interference\cite{zhang1}.

Quantum holography uses a two-photon entangled source and two-photon
coincidence measurement\cite{teich1}. A general two-photon entangled
state can be written as $|\Psi\rangle=\int dx_1 dx_2
C(x_1,x_2)a^{\dag}_s(x_1)a^{\dag}_i(x_2)|0\rangle$, where
$a^{\dag}_j (j=s,i)$ are the photon creation operators for the two
SPDC modes.
$C(x_1,x_2)\sim\langle0|E^{(+)}_{s0}(x_1)E^{(+)}_{i0}(x_2)|\Psi\rangle$
characterizes the two-photon wavepacket for the field operators
$E^{(+)}_{s0}$ and $E^{(+)}_{i0}$ in the source plane. As shown in
Fig. 1(b), while one signal photon passes through a holographic
interferometer and the idler photon travels freely, the evolution of
the field operator is given by Eq. (\ref{1}) (with subscripts
$j=so,sr,i$ instead of $j=o,r$). The signal field is divided into
two parts, $E_{so}^{(+)}$ and $E_{sr}^{(+)}$, serving as the object
and reference waves, respectively. The corresponding IRFs have been
shown in Eq. (\ref{2})(with subscript $so$ instead of $o$ in Eq.
(\ref{2a}) and subscripts $sr$ and $i$ instead of $r$ in Eq.
(\ref{2b})).

Let $E^{(+)}_{s}(x)$ and $E^{(+)}_{i}(x)$ be the field operators of
the signal and idler beams in the detector planes, respectively, the
two-photon wavepacket in the observation planes has the form of
$\langle0|E^{(+)}_{s}(x_1)E^{(+)}_{i}(x_2)|\Psi\rangle$. The
two-photon coincidence counting rate is $R(x_1,x_2)\propto\langle
E^{(-)}_{i}(x_2)E^{(-)}_{s}(x_1)E^{(+)}_{s}(x_1)E^{(+)}_{i}(x_2)\rangle
=|\langle0|E^{(+)}_{s}(x_1)E^{(+)}_{i}(x_2)|\Psi\rangle|^2$. Because
of $E^{(+)}_{s}=E^{(+)}_{so}+E^{(+)}_{sr}$, the rate consists of
four parts: two parts are the two-photon intensities and the other
two parts are the two-photon interference terms given by
\begin{eqnarray}
&&\langle
E^{(-)}_{i}(x_2)E^{(-)}_{sr}(x_1)E^{(+)}_{so}(x_1)E^{(+)}_{i}(x_2)\rangle+c.c.\nonumber\\
&&=\langle\Psi|E^{(-)}_{i}(x_2)E^{(-)}_{sr}(x_1)|0\rangle\times\langle0|E^{(+)}_{so}(x_1)E^{(+)}_{i}(x_2)|\Psi\rangle+c.c.,\label{7}
\end{eqnarray}
which may include the holographic information. Note that this term
defines the spatial interference of two two-photon amplitudes and it
is not involved in ghost interference and ghost imaging.

The two-photon wavepacket of Eq. (\ref{7}) can be calculated by
\begin{equation}
\langle0|E^{(+)}_{j}(x_1)E^{(+)}_{i}(x_2)|\Psi\rangle \propto \int
dx'_0 dx''_0 h_j(x_1,x'_0)h_i(x_2,x''_0) C(x'_0,x''_0), (j=so, sr),
\label{8}
\end{equation}
where $h_j$ is the IRF for beam $j=so,sr$. Particularly, $h_{so}$ is
given by Eq. (\ref{2a}) with the corresponding distances $z_{so}$,
$z_{so1}$, and $z_{so2}$ to replace $z_{o}$, $z_{o1}$, and $z_{o2}$,
respectively; $h_{sr}=H(x,x_0,z_{sr})$ and $h_{i}=H(x,x_0,z_i)$,
where $H()$ is defined by Eq. (\ref{2b}) and $z_{sr}$ and $z_i$ are
the free traveling distances between source and detectors for the
signal and idler beams, respectively. For simplicity, we consider an
ideal two-photon entangled state at the source, satisfying
$C(x'_0,x''_0)=\delta(x'_0-x''_0)$. Equation (\ref{8}) yields
\begin{equation}
\langle0|E^{(+)}_{j}(x_1)E^{(+)}_{i}(x_2)|\Psi\rangle \propto \int
dx_0 h_j(x_1,x_0)h_i(x_2,x_0), (j=so, sr). \label{9}
\end{equation}
This means the fact that the diffraction of the two-photon
wavepacket is equivalent to the diffraction of one-photon which
travels sequently through two paths with IRFs $h_j$ and $h_i$. We
thus obtain $\langle0|E^{(+)}_{so}(x_0)E^{(+)}_{i}(x)|\Psi\rangle$
described by Eq. (\ref{2a}) with $z_{so1}$, $z_{so2}+z_i$, and
$z_{so}+z_i$ to replace $z_{o1}$, $z_{o2}$, and $z_{o}$,
respectively. Also, it has
$\langle0|E^{(+)}_{sr}(x_0)E^{(+)}_{i}(x)|\Psi\rangle\propto
H(x,x_0,z_i+z_{sr})$. An equivalent diagram is shown in Fig. 1(b),
where one of the detectors in the two-photon coincidence measurement
can act as a source. Therefore the two-photon holography can be
easily understood in terms of one-photon case.

We first inspect the proposal by Abouraddy et al \cite{teich1},
where the bucket detection is employed for the beam passing through
the interferometer. The coincidence counting rate in the bucket
detection is $R_{bd}(x)=\int R(x,x_0)dx_0$. For the two-photon
interference term, the integration of Eq. (\ref{7}) gives the
similar form of Eq. (\ref{6}) with $z_o=z_i+z_{so}$,
$z_r=z_i+z_{sr}$, and
$Z'=(z_i+z_{so2})(z_i+z_{sr}-z_{so1})/(z_{sr}-z_{so})$. As a result,
the bucket detector behaves as a spatially incoherent source in the
equivalent diagram. Again, this scheme requires a certain difference
between the object and reference paths. On the other hand, the
longitudinal coherence of the two-photon interferometry is dominated
by the coherence time of the pump beam. The scheme would be
difficult or even impossible when the pump beam has a very limited
coherence time such as a femtosecond pulse laser.

We now consider the point detection regime. According to the
equivalent diagram, the similar result of Eq. (\ref{5}) is obtained
with $z_o=z_i+z_{so}$, $z_r=z_i+z_{sr}$, $z_{o1}=z_{so1}$,
$z_{o2}=z_i+z_{so2}$ and $Z=z_{so1}(z_i+z_{so2})/(z_{i}+z_{so})$. At
the equal path condition, $z_{so}=z_{sr}$, the quadrature phase
factor in the interference term disappears.

Finally, we propose a coherent regime in the two-photon quantum
holography, which can correspond to the plane-wave coherent field
case in the classical holography. The detection system in the signal
beam consists of a lens and a point detector, which is placed at the
foci of the lens. The coherence is due to the fact that all the
encoded photons to be detected have the same momentum. We have
proved that the two-photon interference term (\ref{7}) is the same
as Eq.(\ref{4}) with $z_{o2}=z_i+z_{so2}$\cite{sm}.

In this work, we employ the point detection regime to accomplish
quantum holographic imaging. The experimental setup is shown in Fig.
2. The entangled photon pairs are produced from SPDC in a $5\times
5\times 2\,\mathrm{mm}^3$ beta-barium-borate(BBO) crystal cut for
type-I phase matching. The crystal is pumped by the second harmonic
of a Ti:sapphire femtosecond laser (Mira-900 Coherent Inc.) with
center wavelength $400\,\mathrm{nm}$, and repetition rate
$76\,\mathrm{MHz}$. One of the down-converted beams, named the
signal beam, passes through the interferometer where an object is
set in the object arm, and then reaches detector $D_1$. The other
down-converted beam, the idler beam, travels freely to detector
$D_2$. Both the signal and idler photons are spectrally filtered by
the interference filters of 10 nm bandwidth centered at 800 nm
before arriving the single-photon detectors (Perkin-Elmer
SPCMAQR-14). A time window of 4 ns is chosen to capture the
coincidence counting.

Since a femtosecond pulse as the pump beam has very short coherence
time (120fs), corresponding to the longitudinal coherence length of
$36\mathrm{\mu m}$, we must use the equal-path interferometry. As a
proof-of-principle experiment, the object to be holographically
imaged is an amplitude grating of slit width $b=200\,\mathrm{\mu m}$
and period $d=400\,\mathrm{\mu m}$, described by
$T(x)=\sum_{n=-\infty}^{\infty}\mathrm{rect}[(x-nd-d/2)/b]$, where
$\mathrm{rect(u)}$ is 1 for $|u|\leq 1/2$ and 0 for other values. In
the near-field diffraction, the periodic object can be self-imaged
at a certain distance (Talbot effect), characterized by the Talbot
length $z_T=2d^2/\lambda=40\,\mathrm{cm}$ for
$\lambda=800\,\mathrm{nm}$\cite{song}. So we can definitely know
what we see in the holographic record.

For comparison, we first recast the one-photon holographic imaging
experiment. In this scheme, the signal photon illuminates the
holographic interferometer and is recorded by a scanning detector
$D_1$ while the idler photon is employed as a trigger. To improve
the spatial coherence, a single-slit aperture of width
$100\,\mathrm{\mu m}$ is inserted in the signal beam. The grating is
placed in the object arm of the interferometer at the same distance
of $z_{o1}=z_{o2}=40\,\mathrm{cm}$ to the single-slit and detector
$D_1$. According to Eq. (\ref{5}), the effective diffraction length
is $Z=20\,\mathrm{cm}$ (the half Talbot length), and the image
magnification is two. So this will bring about the self-image of
$T[(x-d)/2]$.

The experimental results are shown in Fig. 3. We first block the
reference arm in the interferometer, and it comes back to the
conventional Talbot self-imaging. We observe the self-image of the
grating in Fig. 3(a), $|T[(x-d)/2]|^2$, which is phase-independent.
Then we release the block to perform the holographic imaging, and
the pattern $T[(x-d)/2]\cos\theta$ is phase-dependent, where phase
$\theta$ is sensitive to the path difference $z_{so}-z_{sr}$. The
in-phase image and out-of-phase image of the grating appear in Figs.
3(b) and 3(c), respectively, by adjusting the path difference
carefully. If the single-slit aperture is taken away in the above
two cases, the image patterns disappear as shown in Figs. 3(d) and
3(e).

Figures 3(d) and 3(e) tell us that the signal beam itself cannot
accomplish the holographic imaging without the help of the single
slit. We now turn to the two-photon nonlocal holographic imaging by
rearrangement of the experimental setup in Fig. 2. Since the pump
beam has poor temporal coherence, the equal-path condition must be
applied to the two-photon interferometry. As has pointed out above,
we must use the point detection scheme in two-photon quantum
holography. The grating is placed at a distance
$z_{so1}=40\,\mathrm{cm}$ from detector $D_1$ and
$z_{so2}=15\,\mathrm{cm}$ from BBO crystal. The distance from BBO
crystal to detector $D_{2}$ is $z_{i}=25\,\mathrm{cm}$. Hence the
effective diffraction length and the image magnification are the
same as the one-photon case. In order to display the nonlocal
feature in quantum holography, detector $D_2$ is scanned across the
beam while $D_1$ is fixed in the two-photon coincidence measurement.

The experimental results in the two-photon holography are presented
in Fig. 4. Again, Fig. 4(a) shows the self-image of the grating when
the reference arm of the interferometer is blocked. This is the
two-photon Talbot self-imaging in the ghost interference scheme,
reported recently by Song et al\cite{song}. In this case, the
interference-diffraction pattern is phase-independent. When the
block is moved away, the in-phase image and out-of-phase image are
shown in Figs. 4(b) and 4(c), respectively. Obviously, these
patterns in Fig. 4 match better with the corresponding ones in Fig.
3 for the one-photon case. If we use a bucket detector to replace
the point detector in the signal photon detection, both in-phase and
out-of-phase image patterns disappear, as shown in Figs. 4(d) and
4(e).

In summary, we have demonstrated experimentally the spatial
interference effect of two two-photon amplitudes given by Eq.
(\ref{7}), which is the origin of quantum holography. We have
analyzed three schemes of quantum holography using a two-photon
entangled source: the point detection, the coherent detection and
the bucket detection. The first two are appropriate for the
equal-path configuration while the last, on contrary, must sustain a
certain optical path difference in the interferometry. Our
experiment has demonstrated the two-photon quantum holographic
imaging in the point detection regime through the two photon
correlation measurement, although the individual detection of the
signal and idler photons do not show any interference pattern. To
make a true hologram, however, it needs to develop two-photon
recording material. Similar to ghost interference and ghost imaging,
the quantum holography reveals nonlocality of quantum entanglement.
Ghost interference and ghost imaging testify the EPR nonlocal
correlation in momentum and in position,
respectively\cite{shih1,shih2}. In quantum holography, however, a
pair of non-commutable physical quantities, the amplitude and phase
of the field, can be nonlocally measured through the two-photon
entanglement. Therefore our experiment on quantum holography may
provide a more authentic version to understand EPR paradox.

This work was supported by the National Natural Science Foundation
of China, Project Nos. 11174038, 11047004 and 10825416, and the
National High Technology Research and Development Program of China,
Project Grant No. 2011AA120102, and the Fundamental Research Funds
for the Central Universities.

Figure captions:

Fig. 1 Sketches of (a) one-photon classical holography and (b)
two-photon quantum holography. BS and M are beamsplitter and mirror,
respectively. In (a), CS, PS, and InCS are the plane-wave coherent
source, point source, and spatially incoherent source, respectively.
RM is the recording material. In (b), CD, PD, BD are the coherent
detection, point detection, and bucket detection, respectively.

Fig. 2 Experimental setup of two-photon holographic imaging. Two
beamsplitters, BS$_1$ and BS$_2$, and two mirrors, M$_1$ and M$_2$,
form an interferometer. NF is the neutral-density filter, and D$_1$
and D$_2$ are two detectors.

Fig. 3 Experimental results of one-photon holographic imaging. CC is
the coincidence counting when detector D$_1$ is scanned and detector
D$_2$ is as a trigger. (a) the self-image of the grating when the
reference path of the interferometer is blocked; (b) and (c) are
respectively the in-phase and out-of-phase images when the reference
path of the interferometer is opened. When the single-slit aperture
in the signal beam is taken away in the cases of (b) and (c), the
self-images disappear as shown in (d) and (e), respectively.

Fig. 4 Experimental results of two-photon holographic imaging. CC is
the coincidence counting when detector D$_2$ is scanned and detector
D$_1$ is fixed. (a)-(c) the same as in Fig. 3. When the bucket
detector is employed in D$_1$ in the cases of (b) and (c), the
self-images disappear as shown in (d) and (e), respectively.
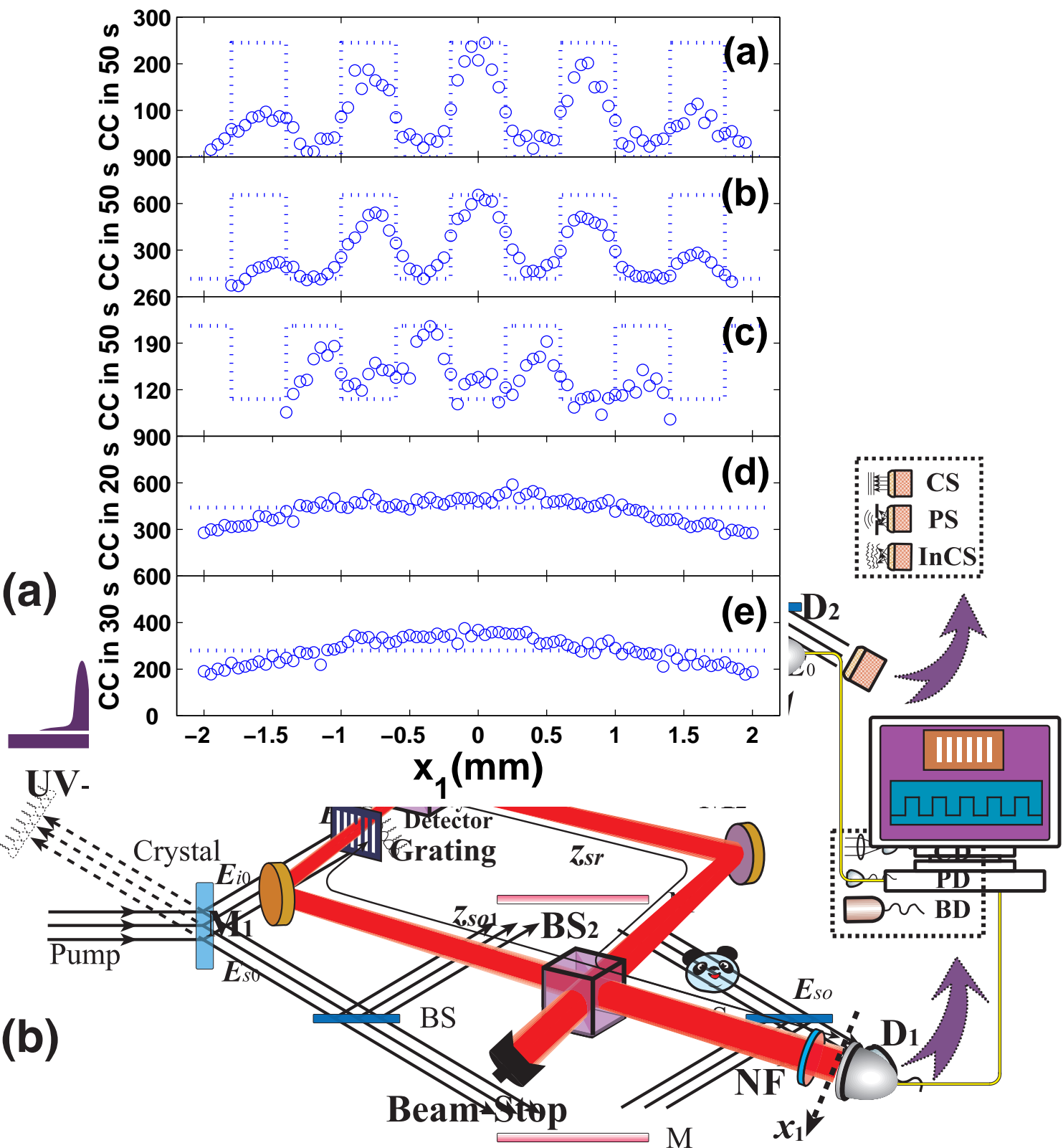
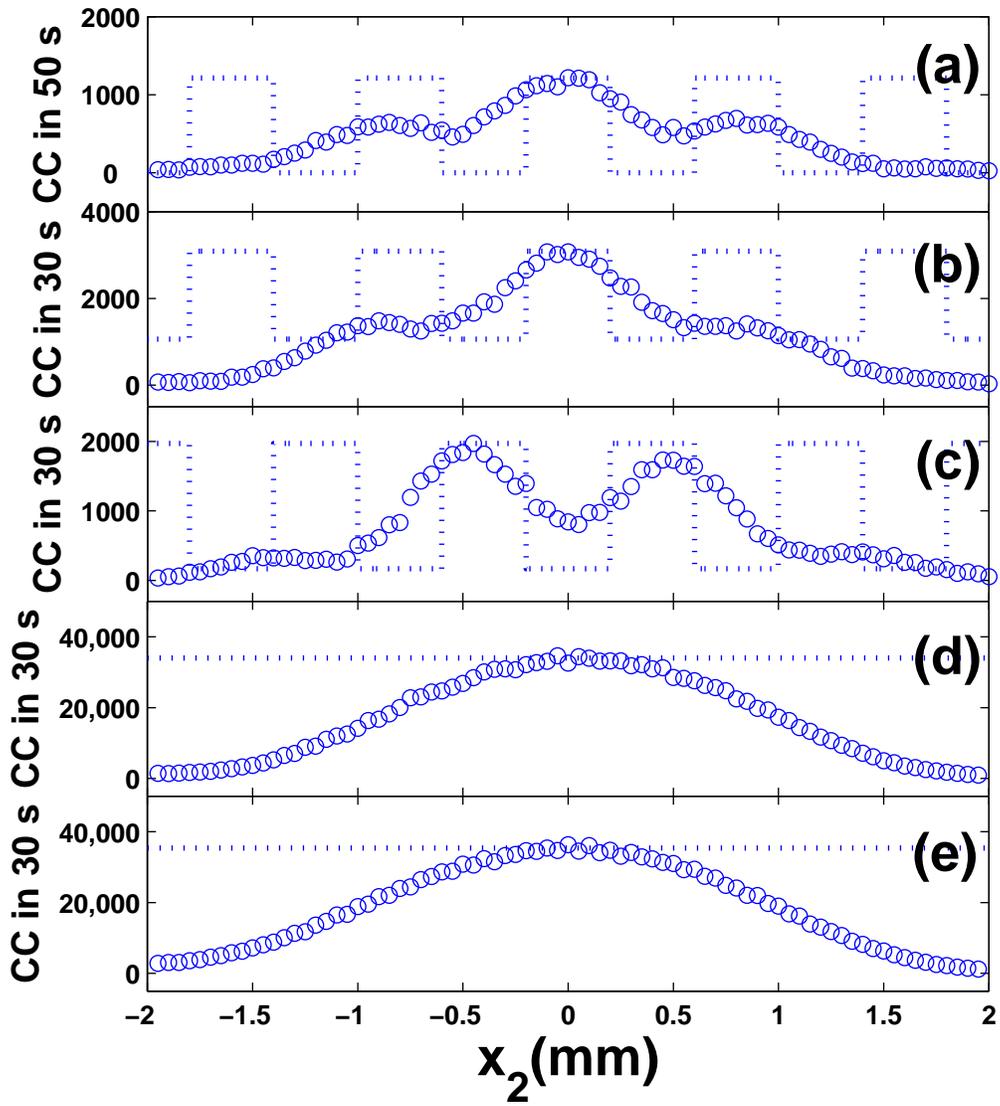
\end{document}